# Understanding Homogeneous Nucleation in Solidification of Aluminum by Molecular Dynamics Simulations

Avik Mahata[a], Mohsen Asle Zaeem[a*] and Michael I. Baskes[b,c,d]

[a] Department of Materials Science and Engineering, Missouri University of Science and Technology, Rolla, MO 65409, USA

[b] Department of Mechanical and Aerospace Engineering, University of California-San Diego, La Jolla, CA 92093, USA

[c] Department of Aerospace Engineering, Mississippi State University, Mississippi State, MS 39762, USA

[d] Los Alamos National Laboratory, NM 87544, USA

**Abstract:**

Homogeneous nucleation from aluminum (Al) melt was investigated by million-atom molecular dynamics (MD) simulations utilizing the second nearest neighbor modified embedded atom method (MEAM) potentials. The natural spontaneous homogenous nucleation from the Al melt was produced without any influence of pressure, free surface effects and impurities. Initially isothermal crystal nucleation from undercooled melt was studied at different constant temperatures, and later superheated Al melt was quenched with different cooling rates. The crystal structure of nuclei, critical nucleus size, critical temperature for homogenous nucleation, induction time, and nucleation rate were determined. The quenching simulations clearly revealed three temperature regimes: sub-critical nucleation, super-critical nucleation, and solid-state grain growth regimes. The main crystalline phase was identified as face-centered cubic (fcc), but a hexagonal close-packed (hcp) and an amorphous solid phase were also detected. The hcp phase was created due to the formation of stacking faults during solidification of Al melt. By slowing down the cooling rate, the volume fraction of hcp and amorphous phases decreased. After the box was completely solid, grain growth was simulated and the grain growth exponent was determined for different annealing temperatures.

[*]Corresponding author; email: zaeem@mst.edu (M. Asle Zaeem)

# 1. Introduction

In metal manufacturing processes involving solidification (e.g., casting [1], welding [2], and laser additive manufacturing [3]), the crystal nucleation from the melt controls the formation and growth of nano- and micro-structures of metals. The solidification structures of materials significantly influence their mechanical and physical properties. If large undercooling can be achieved before crystal nucleation occurs (as in rapid solidification), different and potentially useful forms of crystalline metals may be produced [4]. To predict and control the solidification nano- and micro-structures in different manufacturing processes, a fundamental understanding of mechanisms of crystal nucleation and solidification is necessary.

The crystallization process during liquid to solid transformation can be monitored by using X-ray scattering [5, 6], dilatometry [7], differential scanning calorimetry [8], or microscopic methods [9-11]. But there are several factors that limit the experimental studies of the nucleation process during solidification or crystallization, especially in pure materials (homogenous nucleation) [4]. There are difficulties in quantifying the surface free energy of liquid-solid interfaces and their anisotropy [12]. Also experiments are typically performed at temperatures that differ by hundreds of degrees from the actual nucleation conditions [4]. As a result, experimental studies of nucleation rates in crystallization from the melt cannot provide reliable tests of the classical nucleation theory (CNT) [13, 14]. Another fundamental problem with homogenous nucleation experiments, especially for metallic materials, is that it is difficult to purify a liquid to exclude all the impurities that can catalyze nucleation [15, 16].

Homogenous nucleation from metallic melts is a very complex phenomenon. It starts from the interior parts of an undercooled liquid, and due to the opaque nature of metallic melts, it is very difficult to experimentally detect the nuclei [12, 17]. Therefore, alternative theoretical or computational methods can be used to study homogenous nucleation in pure metals. The problem of nucleation from melt has been studied utilizing different approaches, including theoretical studies based on CNT [13, 14], density function theory (DFT) calculations [18], solid-liquid coexistence by molecular dynamics (MD) simulations [17], and other simulation studies based on phase-field [14], front tracking [19], cellular automata [20], and Monte-Carlo (MC) [21, 22] methods.

Studies utilizing electronic structure calculations based on DFT are limited to few hundreds of atoms and this limits investigation of formation of physical nuclei, which could become large clusters of atoms. In some other computational methodologies such as phase-field, cellular automata and front tracking methods, the length scale is microscale which limits a fundamental understanding of the nucleation process at atomic level.

MD simulations can bridge the gap between the electronic and micro scale computational studies of nucleation and solidification from the melt. MD simulations act between the length scales of DFT and microscale studies, and with the recent advancements in supercomputing, it is now possible to run multi-million atom MD simulations to study phenomena occurring in several hundred nanometer systems. The reliability of MD simulation results significantly depends on the interatomic potentials. DFT calculations alongside experimental data are often used in developing semi-empirical interatomic potentials [23], and MD simulations results are frequently used to provided necessary input information for higher scale models like phase-field models [24-26].

There are few works on homogeneous nucleation during liquid-solid transformation [27, 28] and liquid-vapor transformation [29] by MD simulations. Yasuoka et al. [29] investigated the dynamics of vapor phase homogeneous nucleation in a water system; their predicted nucleation rate was three orders of magnitude smaller than that of the CNT. In metals, it is not straightforward to observe the homogeneous nucleation and solidification processes at the atomistic scale. Shibuta et al. [27] utilized MD simulations and linked the empirical interpretation in metallurgy with the atomistic behavior of nucleation and solidification in pure iron (Fe). These major drawbacks of these works are the use of Finnis-Sinclair (FS) potential [30] and use of isothermal process for all the simulations. Utilizing an isothermal process in MD simulations does not resemble the experimental solidification process. In experiments with slow or fast cooling (quenching), the temperature change will affect the crystal nucleation and solidification processes. The utilized FS potential predicts the melting point of Fe to be 2,400 K, which is much higher than the experimental melting point of Fe (~1,811 K), and consequently results in inaccurate prediction of solid-liquid co-existence properties.

To reliably study the crystal nucleation process from melt by MD simulations, the interatomic potentials used for MD simulations of solidification need to accurately predict the behavior of solid-liquid interfaces. In the early interatomic potentials, which were developed and used for MD simulations of Al such as Lennard-Jones (LJ) [31] and hard-sphere [32] models, only pair interactions of atoms were considered without including the effects of neighboring atoms. Pair potentials do not have environmental dependence (e.g., an atom in the bulk is not similar to an atom on the surface or near a defect site). In reality, the strength of the "individual bonds" should decrease or increase with the change in the local environment during the simulation. Pair potentials do not account for the directional nature of the bond. These are the reasons why pair potentials are not good for predicting the nonlinear phenomena such as failure, plasticity, solidification, melting etc. More complex interatomic potentials were developed for metals to address the shortcomings of pair potentials. Finnis–Sinclair [33] and embedded-atom method (EAM) [34] potentials were developed and used to predict mechanical and physical properties of Al. EAM is a semi-empirical many body potential for the atomistic simulations of metallic systems [35]. FS and EAM predict various properties of several metallic materials and alloys accurately. MEAM interatomic potentials were introduced later to include the directionality of bonding in covalent materials in the EAM and FS formalisms which make the property predications more accurate [36, 37].

Currently the MEAM potentials are widely used in the computational materials science and engineering community to simulate unary, binary, ternary and multi-component metallic systems with different nanostructural features, such as grain boundaries, defects, free surfaces, etc. [38, 39]. In our previous works, we demonstrated the capability of 2NN MEAM potentials in predicting solid-liquid coexistence properties of Fe [40, 41], Ni, Cu, Al [23], and Mg [42], such as melting point, latent heat, expansion in melting, liquid structure factor, and solid–liquid interface free energy and anisotropy. 2NN MEAM potential can also reliably predict room-temperature properties, such as elastic constants, surface energies, vacancy formation energy, and stacking fault energy. The detailed formalism of MEAM and 2NN MEAM can be found in works of Baskes et al. [36] and Lee et al. [43].

To the best of our knowledge, there has been only one experimental study on homogenous crystal nucleation from pure Al melt based on the free boundary (also called the CNT method) and interacting boundary models [44]; the incubation period (or the induction time) and small nuclei were undetectable in this study. There is only one work which used MD simulations [28] to study solidification of Al. However this study doesn't provide quantitative analysis on nucleation, critical nucleus formation, induction time, comparison of MD results to CNT, or details on solid state grain growth.

In this work, we studied the homogenous crystal nucleation from Al melt by MD simulations utilizing the second nearest-neighbor modified embedded atomic method (2NN MEAM) interatomic potential of Al [43]. Homogenous nucleation from Al melt was studied in both isothermal and quench processes. We also provide quantitative details of critical nucleus formation, and comparison of MD with CNT. The regimes of the crystallization process during quenching have been identified. In the last section we also provide detailed analysis of the solid-state grain growth mechanism of pure Al after solidification.

## 2. Simulation Details

MD simulations of homogenous nucleation from pure Al melt were performed in a simulation box with size of 25×25×25 nm$^3$ (64×64×64 unit cells, with 1,000,188 atoms) and with the isothermal-isobaric (NPT) ensemble. Time step of 3 fs was used for all simulations. Temperature and pressure were controlled by Nose-Hoover thermostat and Parrinello-Rahman barostat [45] respectively. Periodic boundary conditions were employed in all three directions. All the MD simulations were performed in LAMMPS [46]. The 2NN MEAM interatomic potential of Al developed by Lee and Baskes [43] was used is this work; we recently tested this interatomic potential which showed accurate predication of solid-liquid coexistence properties of Al [23].

The OVITO visualization package was used to monitor the nucleation and solidification processes [47]. Within OVITO, common neighbor analysis (CNA) was used [48] to identify the local environment of atoms. Using CNA, one can distinguish atoms in different crystal structure regions by calculating the statistics of diagrams formed from the nearest neighbors (NN) of each atom and comparing it with those previously known for standard crystals. For example, if a central atom and its 12 NN form a structure such as fcc, CNA identifies the central atom as fcc. Any such atom is considered an fcc atom. Atoms not identified as fcc, hcp, or any other crystal type implemented in OVITO are identified as amorphous liquid or amorphous solid atoms.

The predicted melting point of Al using a 2NN MEAM MD simulation is 925 K [23], which is in a very good agreement with the experimental value of 934 K. We found that at a temperature close to the melt temperature, the liquid has a fluctuating number of fcc atoms. We wanted to start the nucleation simulations with a pure liquid having no solid regions. In order to find the temperature at which a completely melted simulation box with no fcc crystal can be achieved in a relatively short simulation time (~150 ps), several simulations were performed by increasing the temperature of the simulation box higher than 925 K using 25 K intervals. After 16 intervals, when the temperature reached 1,325 K, we could obtain a completely melted simulation box in ~100 ps. The simulation is continued to 300 ps to make sure the initial melt is properly equilibrated. The CNA of the simulation box for very large time scale is provided in Fig. 1(a). The percentage of amorphous liquid atoms keeps increasing with increasing the annealing temperature. Finally, the box had no crystalline atoms at 1,325 K. The radial distribution function (RDF, g(r)) of the simulation box was calculated for all the temperatures, which is plotted in Fig. 1(b). There are no long-range peaks at 1,325 K. The CNA analysis and RDF plots confirmed that Al was completely melted at 1,325 K.

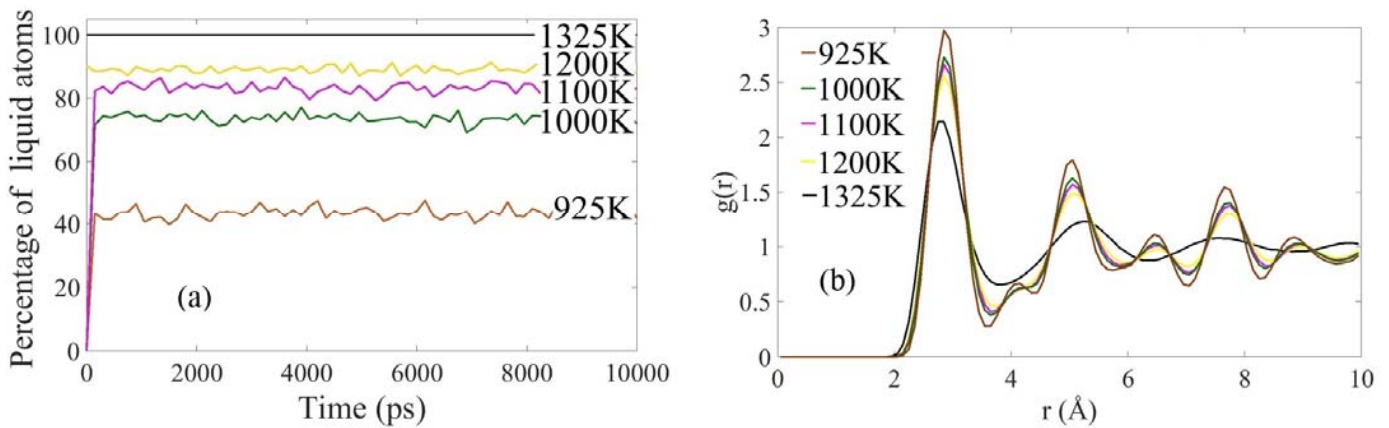

Fig. 1. (a) Percentage of amorphous liquid atoms at different temperatures; (b) The radial distribution function of Al melt in Fig. 1(a) showing liquid characteristic with no long range peak at 1,325 K.

For isothermal simulations, the Al melt was isothermally solidified at temperatures between the range of 300 K and 800 K with 50 K intervals. Maximum nucleation rate was observed to be between 400 K and 500 K, so we ran more simulations with 25 K intervals to more accurately determine the exact critical temperature of nucleation (when the nucleation rate is maximum). Each isothermal simulation was repeated five times to evaluate the possible errors. Each isothermal simulation was run for a total of 500 ps (167,000 time steps) to simulate the crystal nucleation and solidification.

We also performed solidification by quenching with different cooling rates of $5.83 \times 10^{10}$ $Ks^{-1}$, $5.83 \times 10^{11}$ $Ks^{-1}$ and $5.83 \times 10^{12}$ $Ks^{-1}$. Different cooling rates were applied by changing the number of total time steps. The initial temperature of the melt was 1,325 K, then the melt was cooled down to 450 K in 150 ps, 1,500 ps and 15,000 ps (5,000,000 time steps), which resulted in cooling rates of $5.83 \times 10^{12}$ $Ks^{-1}$, $5.83 \times 10^{11}$ $Ks^{-1}$ and $5.83 \times 10^{10}$ $Ks^{-1}$, respectively. 450 K was chosen because it is lower than the critical temperature found in the isothermal process (see section 3.6). The quenching method was used to mimic the actual experimental procedure to produce undercooling where the temperature decreases from above the melting temperature with a certain cooling rate. This method of simulation is closer to what is performed experimentally and differs from the previous MD simulations of homogenous nucleation which usually utilized isothermal simulations [27]. In experiments, cooling rates in rapid solidification of bulk Al lie between $10^4$ and $10^7$ K/s [49-51], notably much slower than the rates used in MD.

## 3. Results and Discussion

### 3.1 Crystal Structure of Nuclei

The primary observation of nucleation in MD simulations shows the formation of nuclei from the melt. CNA and visual inspection are used to study the structure of the nucleus throughout the quenching and annealing. The formation of crystal structures and stacking faults occurred in the same way for both the isothermal and quenching processes. Fig. 2(a) shows that the crystalline nuclei form in different parts of the melt; the unstructured melt is removed from the simulation box so the crystalline nuclei can be seen. The spherical nature of the nuclei is observed visually at first. The atomic coordinates of the cluster atoms in the specified nucleus in Fig. 2(b) are extracted, and the measured distances from the surface atoms of the nucleus to the central atom show almost the same value in all directions. The magnified nucleus in Fig. 2(b) shows atoms with fcc (green) and hcp (red) crystal structures using CNA. The nearest neighbor distance for fcc Al should be 2.86 Å as the lattice constant is 4.05 Å. The distance between two nearest atoms within the fcc (green) atoms in Fig. 2(b) is ~ 2.86 Å. It should be mentioned that few solid amorphous atoms get trapped inside the fcc/hcp crystalline phase nucleus, which don't have enough neighbors to be detected as a solid crystalline phase.

We calculated the difference between formation energies of fcc and hcp Al to be only 0.03 eV, whereas the difference between formation energies of fcc and bcc Al was determined to be 0.12 eV [23]. Since there is a random thermal fluctuation of energy during solidification, this thermal fluctuation of energy can cause formation of hcp stacking faults in the Al system, but it is not enough to promote formation of bcc atoms. By rotating the simulation box in Fig. 2(b) around the <111> direction, the stacking fault plane can be observed (Fig. 2(c)), and the stacking fault is detected to have hcp crystal structure.

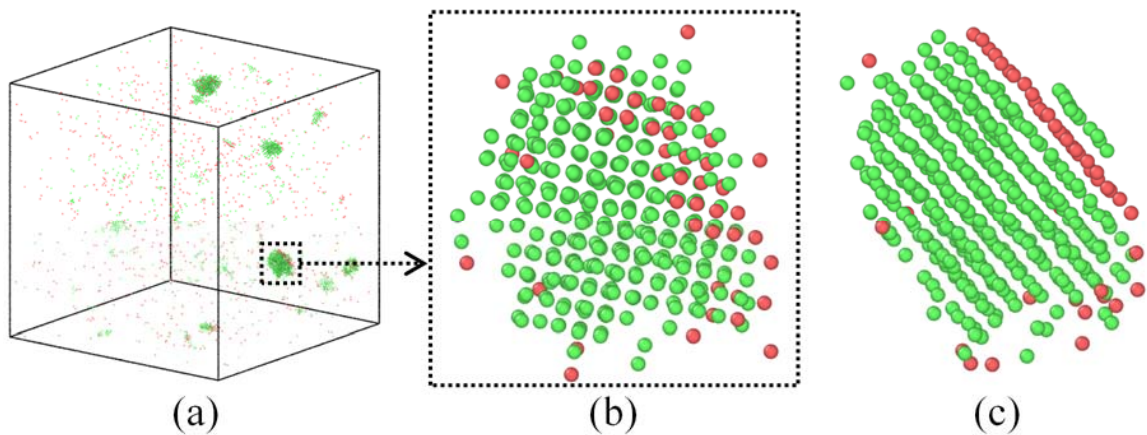

Fig. 2. (a) Formation of nuclei after 1,050 ps of simulation at annealing temperature of 400 K, (b) Magnification of one of the nuclei in Fig. 2(a), and (c) Rotation of the nucleus in Fig. 2(b) around the <111> direction showing the stacking fault plane. FCC atoms are in green and hcp atoms are in red.

3.2  Critical Nucleus Size

The minimum size required for continuous growth of a crystalline nucleus is known as the critical nucleus size. In this study, the size of a nucleus is taken as the average, as discussed below, of the maximum length of the nucleus in x, y and z directions; this is assumed to be equivalent to the diameter of a spherical shaped nucleus. The nucleus size and number of atoms in the nucleus are determined by direct observations. Before a nucleus reaches its critical size, for a short period of time (nucleus origin time, discussed in Section 3.9), the nucleus gains and loses atoms. We assume that a crystalline nucleus reaches its critical size when it doesn't lose any atom back into the liquid. Fig. 3 shows examples of nuclei size and number of atoms in the nuclei versus simulation time for isothermal and quenching cases. The arrows in Fig. 3(a) and (b) show when the nuclei reach the critical size. The number of atoms in the critical nucleus is ~1400 atoms at 700K (Fig. 3(a)) and for the case with a cooling rate of $5.83 \times 10^{11}$ Ks$^{-1}$ in Fig. 3(b), the critical sized nucleus has ~1000 atoms. After reaching the critical size, the crystalline nuclei grow in size and gather more fcc and hcp atoms. The evolution of critical nucleus can also be monitored by potential energy change with time and the visualization snapshots (Inset Fig. 3a-b). The crystalline nuclei reach the critical size slightly before the sudden change of slope. At that point the nucleus has become large enough to overcome the free energy barrier for phase separation. A critical nucleus does not become smaller after it reaches the critical size. The critical nuclei are found to be quiet stable against the mobility of liquid phase, structural change, i.e. fcc-hcp, or continuously changing shape.

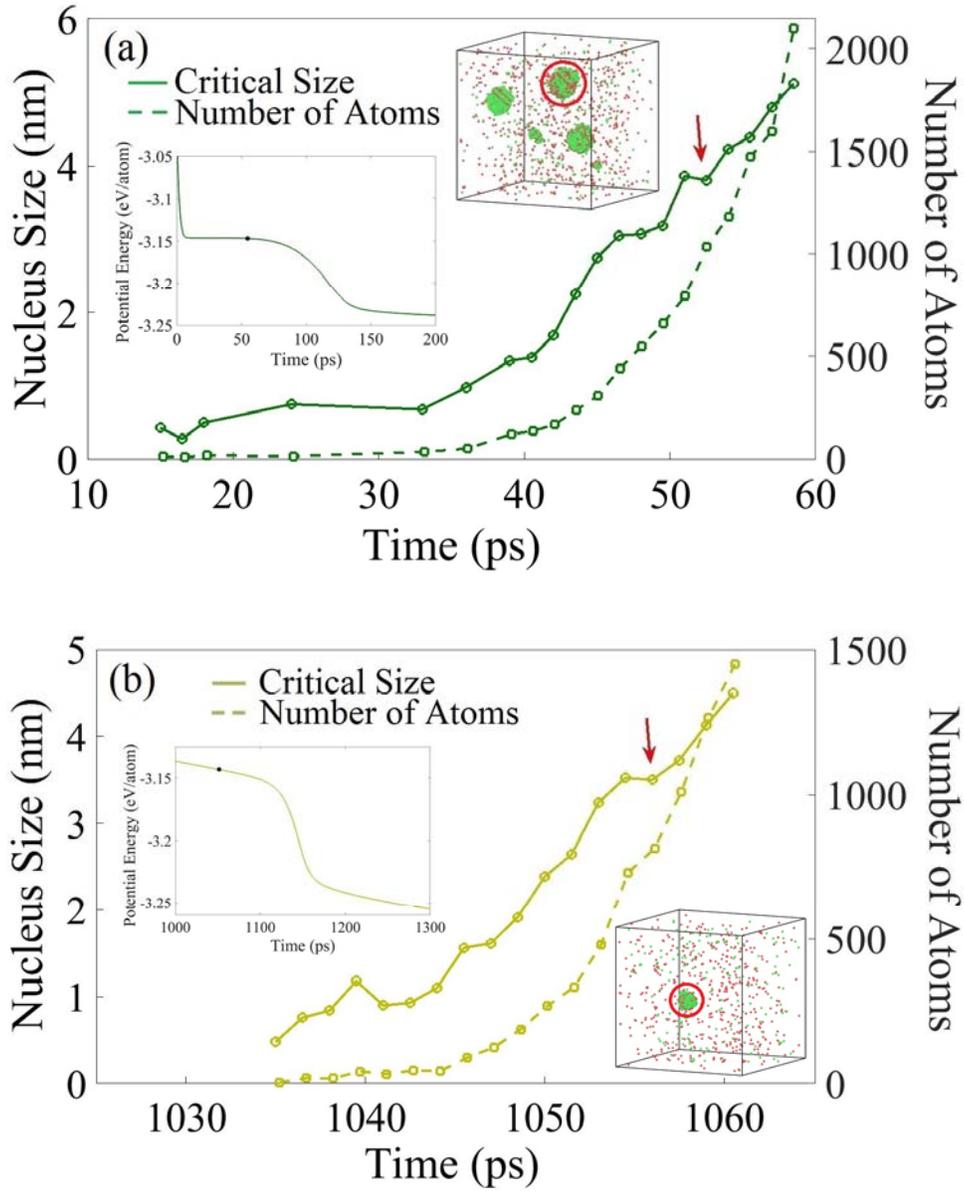

Fig. 3. Nucleus size versus simulation time for: (a) isothermal process at 700 K annealing temperatures, and (b) quenching process at the cooling rate of $5.83 \times 10^{11}$ Ks$^{-1}$. Arrows show the number of atoms in the critical sized nuclei. The inset graph of potential energy vs time shows the change in slope. The exact point of formation of critical nucleus formation is marked by a black dot. The other inset of the simulation box shows the critical nucleus.

As it was mentioned before, each simulation was performed 5 times; after the nuclei reach the critical size, a set of 5 random critical nuclei are chosen at each annealing temperature from each isothermal simulation; total of 25 critical nuclei were selected for each annealing temperature. The average critical size and its standard deviation versus annealing temperature are plotted in Fig. 4(a). The average size of critical nuclei is between ~0.82 nm and ~4 nm (Fig. 4(a)) for all annealing temperatures in the isothermal process. The critical size of crystalline nuclei increases as the annealing temperature increases.

In the quench processes, the average critical size of nuclei is found to be ~1.8 nm for $5.83 \times 10^{12}$ Ks$^{-1}$, ~3.49 nm for $5.83 \times 10^{11}$ Ks$^{-1}$, and ~4.5 nm for $5.83 \times 10^{10}$ Ks$^{-1}$ cooling rate. With a slower cooling rate the nucleation occurs at higher temperatures, which results in a larger critical size for nuclei. It should be noted that nucleation rates decrease with slower cooling rates (discussed in section 3.7), and the nuclei can grow larger before the whole simulation box becomes solid.

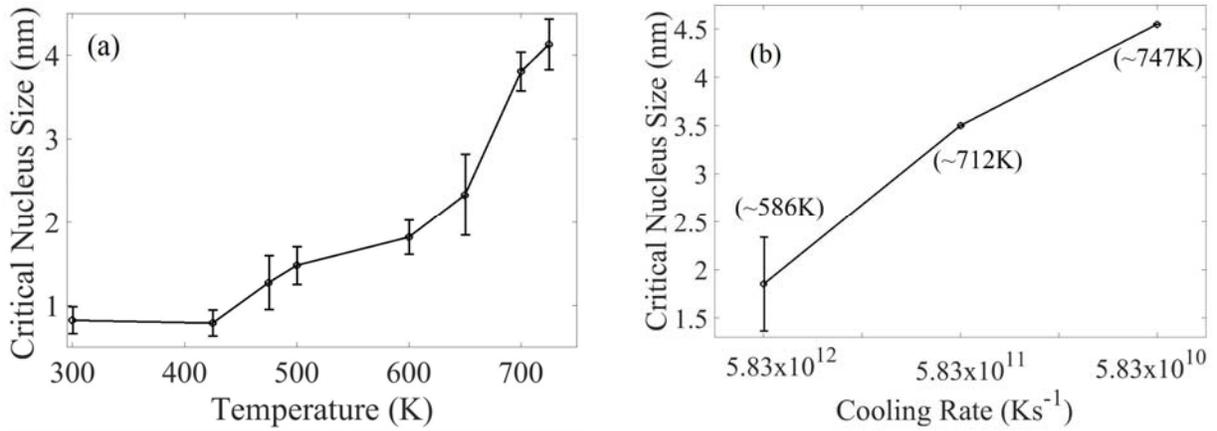

Fig. 4. The critical nuclei size versus (a) annealing temperatures in the isothermal process and (b) different cooling rates. Error bars are shown as the standard deviation for 5 randomly chosen critical nuclei for isothermal simulations. The temperature at which the critical nucleus forms is shown for quenching in Section 3.3 below

During the isothermal process, the maximum number of crystalline nuclei in the system varies (Fig. 5) with annealing temperature. When a relatively low annealing temperature is applied (below 600 K), the nucleation starts instantly, and since the driving force for solidification is very high, fcc crystalline atoms form all over the simulation box in the early stages of simulation. This will result in formation of multiple critical nuclei simultaneously, and a higher number of nuclei will form but they grow to a smaller size compared to the nuclei in higher annealing temperatures. The maximum number of separable nuclei in the simulation box for annealing temperatures between 350 K and 650 K is more than 40 nuclei in 25 nm$^3$. Above 650 K the maximum number of separable nuclei is reduced; for example, at 700 K and 725 K, 12 and 9 nuclei are detected, respectively. While the number of nuclei is reduced by increasing the annealing temperature above 500 K, each nucleus can grow to a much larger size before the simulation box is completely solid. In the quenching process, the maximum number of separable nuclei varies between 9 to 15 for different cooling rates, which is similar to that of 700 K and 725 K isothermal cases. The maximum number of separable nuclei is seen at 715 K, 665 K and 655 K for $5.83 \times 10^{10}$ Ks$^{-1}$, $5.83 \times 10^{11}$ Ks$^{-1}$ and $5.83 \times 10^{12}$ Ks$^{-1}$ quench rates, respectively.

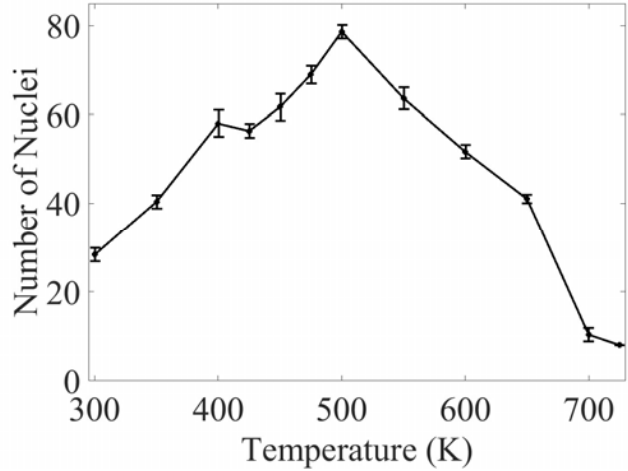

Fig. 5. Maximum number of critical nuclei at different annealing temperatures during the isothermal process. The error bars show standard deviation for 5 replicate simulations at each annealing temperature.

3.3 Temperature dependence of nucleation

The crystal nucleation from an undercooled Al melt predicted by MD simulations is shown in Fig. 6 for two isothermal cases (annealing at 475 K and 700 K), and two quenching cases (cooling rates of $5.83 \times 10^{10}$ Ks$^{-1}$ and $5.83 \times 10^{11}$ Ks$^{-1}$); only fcc Al atoms are shown for a better visualization of nuclei. The instantaneous time and temperatures during each quenching process are also shown in Fig. 6(c) and Fig. 6(d).

As mentioned before, the number of nuclei in the system is reduced with increasing the annealing temperature above 500 K (Fig. 5); the same conclusion can be made by comparing Fig. 6(a) and Fig. 6(b). These figures also show that each nucleus can grow much bigger in size at a higher annealing temperature. Nucleation by quenching in Fig. 6(c) and Fig. 6(d) shows a very similar behavior to the nucleation of isothermal process at 700 K (Fig. 6(b)). This indicates that in the quenching process nucleation starts at high temperatures with a small number of nuclei which can grow in size.

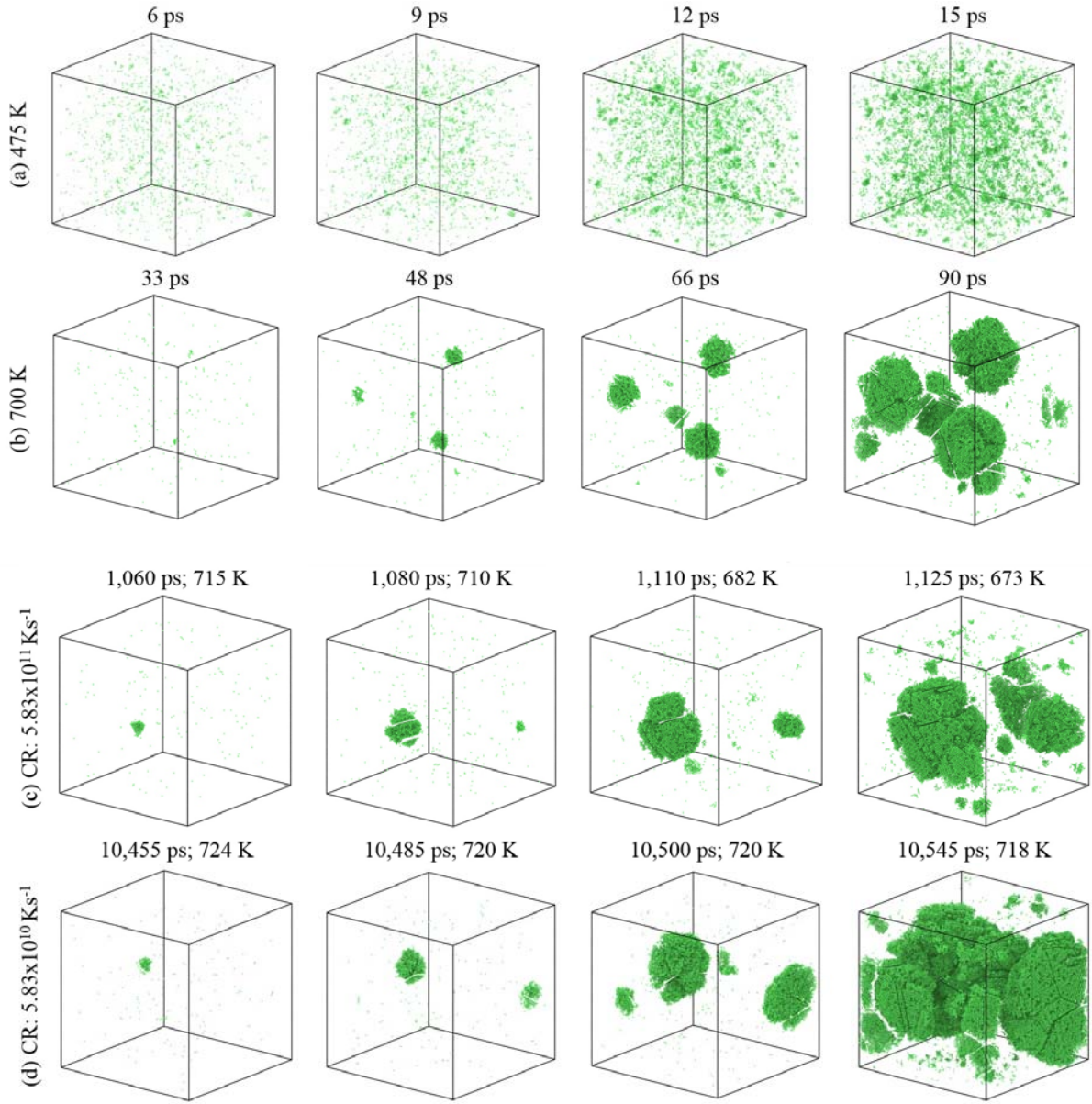

Fig. 6. Snapshots of nuclei formation and growth during solidification for two isothermal processes at annealing temperatures of (a) 475 K and (b) 700 K, and for two quenching processes with cooling rate of (c) $5.83 \times 10^{11}$ Ks$^{-1}$ and (d) $5.83 \times 10^{10}$ Ks$^{-1}$. Fcc are shown with green color; hcp and amorphous atoms are ignored for a better visibility of nuclei.

The instantaneous temperature for crystal nucleation during quenching can be determined by plotting percentage of crystalline atoms versus temperature change (Fig. 7). Fig. 7 shows that during quenching the nucleation process starts between ~747 K and ~712 K for the slower cooling rates of $5.83 \times 10^{10}$ Ks$^{-1}$ (Fig. 7 (b)) and $5.83 \times 10^{11}$ Ks$^{-1}$ (Fig. 7 (c)). For a high cooling rate of $5.83 \times 10^{12}$ Ks$^{-1}$ (Fig. 7 (a)) the nucleation starts below 700 K, and the exact temperature of formation of first nucleus is found to be ~586 K from dumps (per atom data) available from LAMMPS (such as Fig. 6). The number of fcc/hcp atoms is very low for $5.83 \times 10^{12}$ Ks$^{-1}$ cooling rate that it doesn't reflect the nucleation starting temperature in Fig. 7(a). The most solid atoms for this cooling rate remain at amorphous configuration, and the change in overall crystal structure is not significant until about 575 K. The number of nuclei is very high, but they grow much less in size compare to the nuclei in slower cooling rates. Overall the quenching simulations suggest that the nucleation starting temperature is between 586 K -747 K.

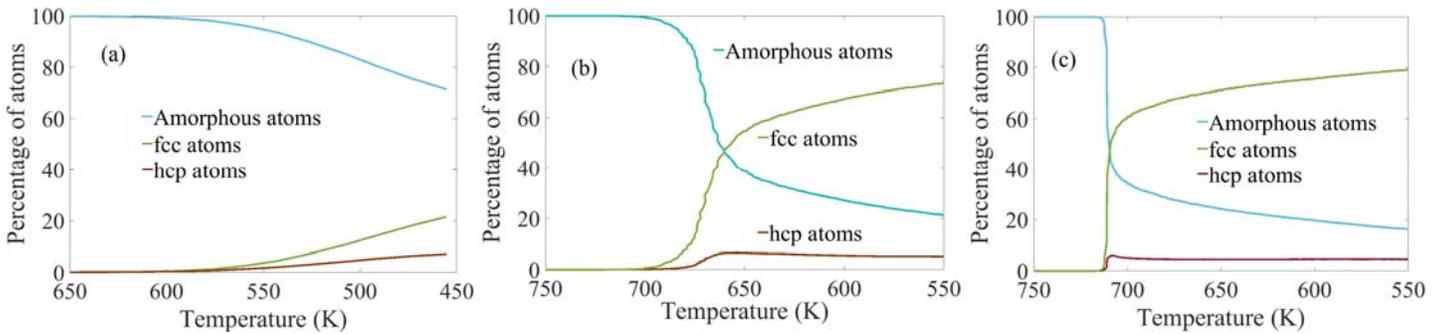

Fig. 7. The change in types of atoms with temperature for three different cooling rates of (a) $5.83 \times 10^{12}$ Ks$^{-1}$, (b) $5.83 \times 10^{11}$ Ks$^{-1}$, and (c) $5.83 \times 10^{10}$ Ks$^{-1}$.

3.4  Crystallization during nucleation

The percentage of atoms having different structures (amorphous, fcc or hcp) is plotted in Fig. 8(a)-(c) at three different annealing temperatures for the isothermal process. The percentage of fcc atoms for 700 and 725 K (~60-65 %) is slightly higher than the percentage of fcc atoms generated at lower temperatures (~50-55 %). At very low temperatures such as 350 K, the percentage of fcc atoms is very low. The lower amount of fcc atoms also causes a very low nucleation rate at 350 K. At lower temperatures (below 600 K), the accumulation of fcc atoms starts immediately. At the same time number of hcp stacking faults grows and number of amorphous atoms decreases. At 725 K the initial nucleation starts later than any other temperature. At temperatures, higher than 725 K, there is no nucleation in 600 ps.

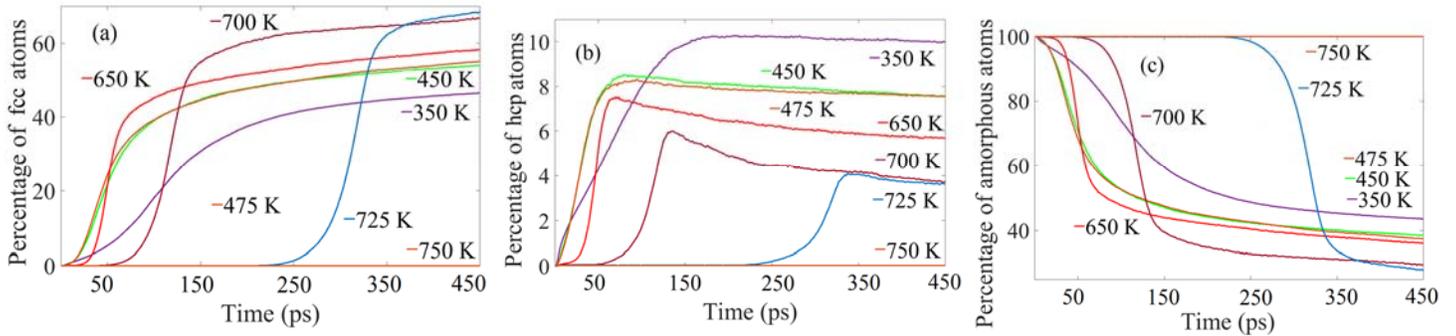

Fig. 8. Percentage of atoms with fcc, hcp and non-structured amorphous configurations during simulations are plotted for different annealing temperatures for (a) fcc atoms (b) bcc atoms and (c) amorphous atoms. Data from each the 1.5 ps were used to generate the figures.

Percentage of atoms having different structures (amorphous, fcc or hcp) for different quench rates are also plotted in Fig. 9(a)-(c). In quenching the process of nucleation starts after first regime when mostly sub-critical nucleus/nuclei forms and dissolves. Slowly the temperature reduces to nucleation regime and then the nucleation happens very fast. However the rate of formation of fcc atoms differs for different cooling rates. The slower the cooling rate is the more time the melt has to solidify the fcc/hcp atoms and this lowers the number of amorphous atoms. With a slower cooling rate number of hcp stacking faults also decreases.

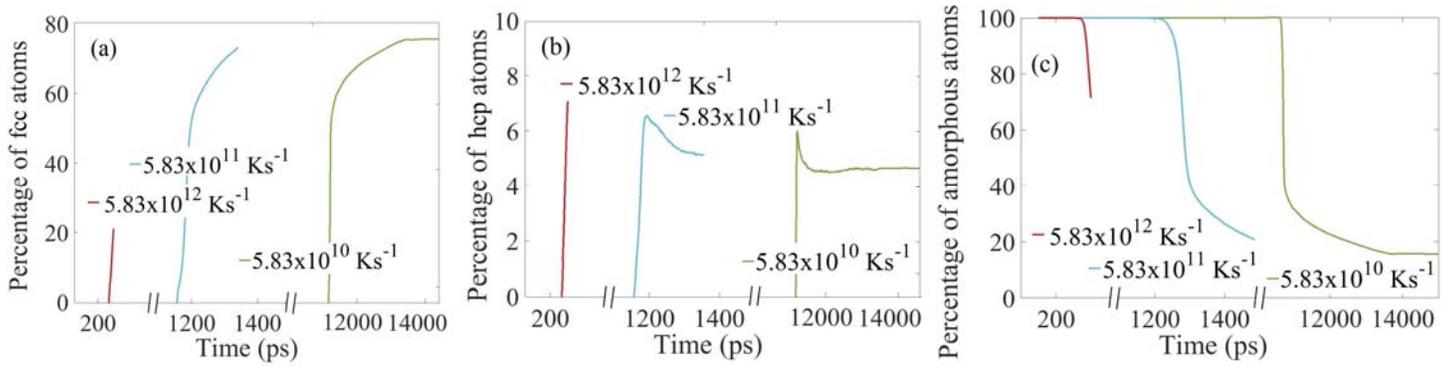

Fig. 9. Percentage of atoms with fcc, hcp and non-structured amorphous configurations during quenching vs. time. Data from all the time-steps were used to generate the figures.

3.5　Temperature-dependent nucleation regimes

As it was shown in the previous section 3.3, nucleation is a temperature driven phenomenon, and a change in temperature affects the rate and behavior of nucleation. Potential energy is one of the fundamental quantities that correlates temperature with nucleation and solidification processes. Fig. 10(a) shows the potential energy versus simulation time for different isothermal annealing temperatures. When the Al melt is brought directly to a low annealing temperature, there is a very sharp drop in the initial potential energy due to the specific heat of the liquid.

Below 600 K, the Al melt starts solidifying immediately within the first few time steps. For higher annealing temperatures (such as 650 K, 700 K and 725 K) the solidification doesn't happen immediately. The time required to form the first critical nucleus (or nuclei) after starting the annealing is ~40 ps, ~75 ps and ~ 250 ps for annealing temperatures of 650 K, 700 K and 725 K, respectively.

In isothermal cases, we can roughly see from Fig. 10(a) that the behavior of melt changes with temperature. At 750 K and above, the potential energy is flat for the entire simulation time indicating that no nucleus forms. Between 600 K and 725 K three distinct regimes can be observed. We initially see a flat region similar to 750 K where no nucleation occurs. Then the curvy decay of the potential line indicates that nucleation happens at this stage. This indication is verified against per atom data (i.e. LAMMPS dumps), and it is found that the decay of potential energy and nuclei formation happen simultaneously. The final flat region shows the start of the solid-state grain growth. From 350 K to 575 K the sharp decrease in potential energy starts immediately and continues until the end of the simulation, which shows that the nucleation starts immediately, and towards the end of the simulation the potential energy curves becomes parallel to each other. At lower annealing temperatures (below 350 K), even though the Gibbs free energy difference between fcc/hcp and amorphous liquid atoms is large, the low mobility of atoms at low temperatures affects the kinetics of nucleation and not all the liquid atoms can form crystalline structures. An amorphous solid structure is retained. At the final stage of all cases, curves become slowly parallel to the time axis with an offset from each other. This offset is due to the specific heat of the solid phases, which results in lower potential energy at lower temperature. Overall, the isothermal simulations do not give the temperature range for the different stages of nucleation; these simulations only indicate that liquid Al melt has to be below 725 K for nucleation to occur.

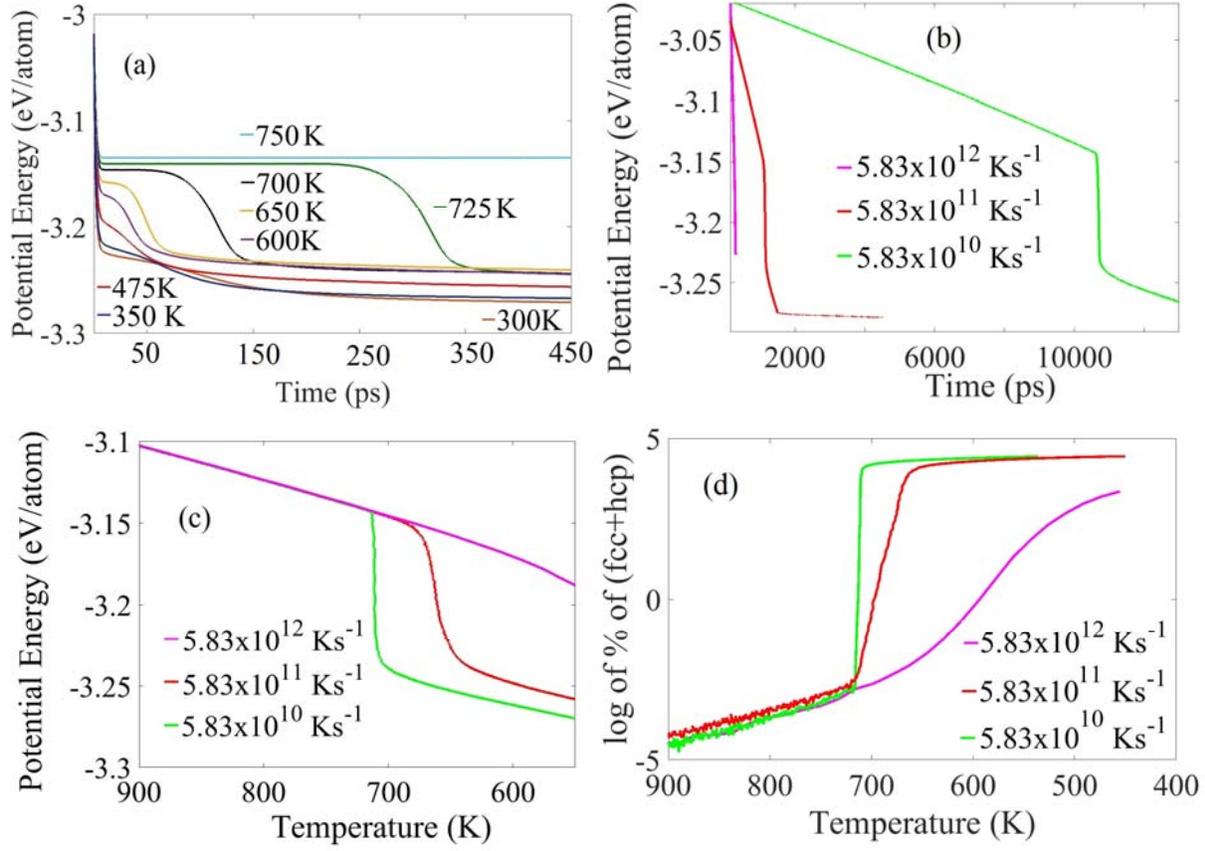

Fig. 10. Potential energy versus time for (a) isothermal process at different annealing temperatures, and (b) quenching process with different cooling rates from 1,325 K to 450 K. (c) Potential energy versus temperature for quenching process with different cooling rates. (d) Log of percentage of fcc/hcp crystalline atoms versus temperature.

Fig. 10(b) shows the potential energy versus time for the quenching process with different cooling rates from 1,325 K to 450 K. It is not possible to identify the nucleation regimes by this figure, but the plot of the potential energy versus temperature, Fig. 10(c), reveals three temperature-dependent regimes very clearly. As shown in Fig. 10(c), the potential energy decreases when temperature decreases during quenching simulations. During this period (> 725 K), sub-critical nucleus/nuclei form and dissolve back into the melt. For the low cooling rates ($5.83 \times 10^{10}$ Ks$^{-1}$ and $5.83 \times 10^{11}$ Ks$^{-1}$), the sharp change in the slope (Fig. 10(b)) indicating a sudden decrease in potential energy occurs at temperatures range of 715-725 K. The beginning of the sharp change in slope shows the start of formation of critical nuclei (at ~725K), and once the first sub-critical nucleus reaches the critical size the crystallization happens very fast at low cooling rates. The super-critical nuclei grow bigger in size until the simulation box is completely solid, and solid-state grain growth starts; this region can be easily identified for low cooling rates (e.g., < 710 K for $5.83 \times 10^{10}$ Ks$^{-1}$). For the low cooling rates, the almost vertical slope line (Fig. 10(c)) signifies the release of latent heat due to crystallization (or solidification). This event represents the fast and spontaneous formation of solid nuclei during solidification. Fig. 10(c) shows nuclei formation can only happen in a temperature range depending on the cooling rate. This finding shows the drawbacks of isothermal simulations and clearly shows the existence of different temperature regions in the solidification process.

Initially atoms attempt to crystallize from the melt by formation of small clusters of fcc and hcp atoms, but as the simulation progresses most of these clusters of atoms dissolve back into the liquid phase. This region can be identified as the sub-critical (unstable) nucleation regime in Fig. 10(d) where there are small fluctuations in the total number of crystalline atoms. The sudden change in the slope shows the regime change from sub-critical to super-critical (stable) nucleation, and the temperature at which this transition occurs is named $T_{sc}$. The exact value of $T_{sc}$ depends on the cooling rate, and remains between ~715-725 K for $5.83 \times 10^{10}$ Ks$^{-1}$ and $5.83 \times 10^{11}$ Ks$^{-1}$. Multiple super-critical nuclei are formed in the system following the formation of the first critical nucleus and they grow until the whole simulation box is solid and solid-state grain growth starts.

The second sudden change in the slope shows another regime change from super-critical nucleation to solid-state grain growth, and the temperature at which this transition occurs is named $T_{gg}$. The difference between $T_{sc}$ and $T_{gg}$, is very small (only 9 K for $5.83 \times 10^{10}$ Ks$^{-1}$ cooling rate, $T_{sc}$ = 724 K start and $T_{gg}$ = 715 K end) for the slowest cooling rate. Super-critical nuclei will grow until the whole box is solid, and then solid-state grain growth occurs. This solid-state grain growth and end of nucleation are the same temperature ($T_{gg}$). The grain growth is not a part of nucleation process, but an essential part of solidification. So overall the solidification process can be divided into three temperature based thermodynamics regimes, i) the sub-critical (unstable) regime, ii) the super-critical (stable) nucleation regime when, multiple critical nuclei form along with the growth of the previously formed critical nuclei, and iii) solid-state grain growth regime.

In any experimental method, the Al melt needs to release heat to go down to a specific annealing temperature. Even if the Al melt can be brought to a constant temperature environment instantly, it is practically impossible that the Al melt will go down to the lower annealing temperature immediately. In other words, the quench rate in isothermal processes are infinite. In quenching we showed, the $T_{sc}$ for 5.83x10$^{10}$ Ks$^{-1}$, 5.83x10$^{11}$ Ks$^{-1}$, 5.83x10$^{12}$ Ks$^{-1}$ are 724 K, 715 K and 586 K respectively. So, we can see as the cooling rate decreases the $T_{sc}$ increases. In MD, the quench rates are very high. If we assume in bulk experiments the cooling rate to be 1-100 Ks$^{-1}$ the nucleation temperature should be higher (above ~725 K). So, in the real world there is no nucleation at all at lower annealing temperature (i.e. 650 K for Al). We also observed the slope in the super-critical region (Fig. 10(d)) is getting more vertical as the cooling rate is reduced. So, in experiments when the cooling rate is much slower, the $T_{sc}$ and $T_{gg}$ can be almost the same temperature.

3.6  Nucleation rate versus annealing temperature: isothermal solidification

It is evident from the previous section that annealing temperature certainly affects the nucleation process, so it is expected that it would also affect the nucleation rate. The nucleation rate for each annealing temperature is calculated by fitting a line to the data on number of nuclei versus time, where the slope of the line is the nucleation rate (Fig. 11). The nucleation rate increases as the annealing temperature increases from 300 K to 475 K (Table 1). At room temperature (300 K) very few separable crystalline nuclei can be found; for higher annealing temperatures, the kinetic energy of atoms increases, which helps liquid atoms overcome the activation or free energy barrier to produce critical sized crystalline nuclei.

The critical temperature of nucleation is the temperature at which the nucleation rate is a maximum. Our initial simulations were done for undercooling temperatures between 300 K and 800 K with an interval of 100 K. We found that the nucleation rate is a maximum between 400 K and 500 K; therefore to find the exact critical nucleation temperature, more simulations were performed between annealing temperatures of 400 K to 500 K with an interval of 25 K. From the slopes of the fitted lines in Fig. 11, the maximum nucleation rate of 5.74x10$^{35}$ m$^{-3}$s$^{-1}$ occurs at the annealing temperature of 475 K (Table 1). The typical nucleation rate for the homogeneous nucleation of a pure metal near the critical temperature has been estimated previously from experiment to be in the order of $10^{30}$ and $10^{40}$ m$^{-3}$ s$^{-1}$ [52], which is comparable to our MD results.

Since the nucleation rate is maximum at 475 K, we can come to a conclusion that ~475 K is the critical temperature of nucleation for Al. The calculated critical temperature from MD is ~$\frac{T_m}{2}$, where $T_m$ is the melting temperature. Once the solidification progresses the distance between different nuclei is reduced, and the simulation box eventually transforms into the bulk solid crystalline Al with hcp solidification defects and grain boundaries.

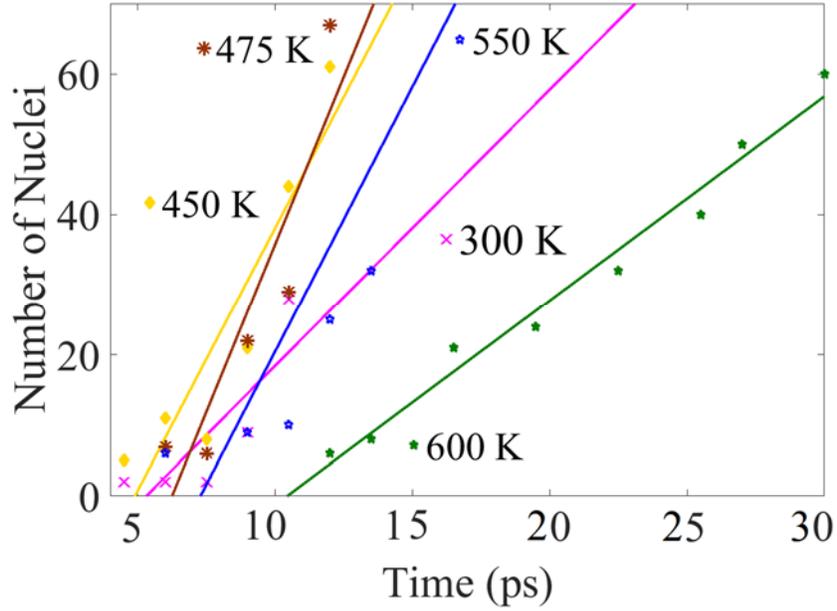

Fig. 11. The number of nuclei as a function of time at various annealing temperatures. The slopes of lines are the nucleation rates at different annealing temperatures which are reported in Table 1. Note that the maximum nucleation rate occurs at 475 K, which defines the critical temperature of nucleation for Al.

Table 1. Nucleation rates at different annealing temperatures. The statistical error is estimated by obtaining the slopes for 5 different simulations of each annealing temperature.

| Temperature ($T$) (K) | 400 | 425 | 450 | 475 | 500 | 550 | 600 | 700 | 725 |
|---|---|---|---|---|---|---|---|---|---|
| Nucleation rate ($I$) ($10^{35}$ m$^{-3}$s$^{-1}$) | 4.00±0.13 | 4.00±0.10 | 4.48±0.08 | 5.74±0.07 | 5.32±0.05 | 4.54±0.06 | 3.51±0.01 | 0.07±0.00 | 0.03±0.00 |

## 3.7 Nucleation rate versus cooling rate: solidification by quenching

In quenching crystallization begins by formation of small clusters of atoms at high temperatures. Many of them form and dissolve back into the liquid; a few will survive. The nuclei starts forming after sometime. The beginning time for the nuclei depends on the cooling rate, slower the cooling rate the later the nucleus (nuclei) forms. From Fig. 12 and Table. 2, the nucleation rates are obtained in the same way it was obtained for the isothermal process. The time in Fig. 12 is a small part of time among the whole time steps. Most of the nucleation happens between this part, so it is chosen to study the nucleation rate in quenching. It is shown in Table. 2 that the nucleation rate goes down from $1.12 \times 10^{35}$ $m^{-1}s^{-1}$ at cooling rate of $5.83 \times 10^{12}$ $Ks^{-1}$ to $8 \times 10^{33}$ $m^{-1}s^{-1}$ at cooling rate of $5.83 \times 10^{10}$ $Ks^{-1}$.

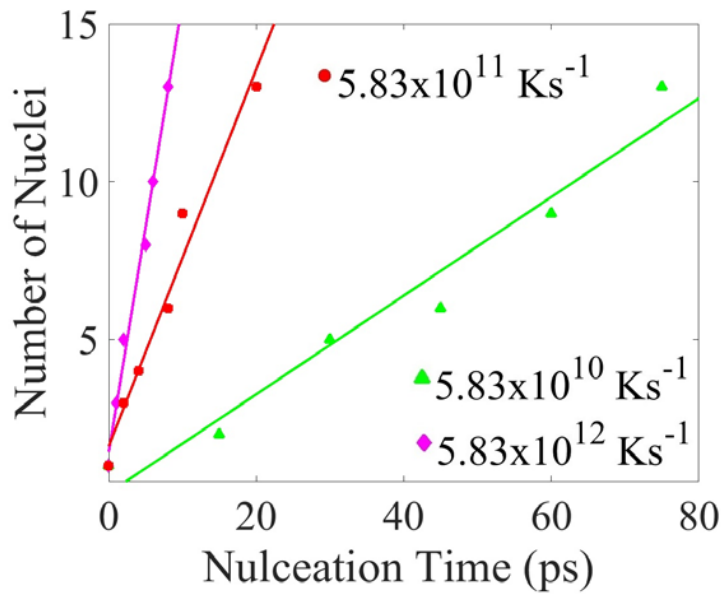

Fig. 12. The number of nuclei as a function of time for various quench rates. The slope of these curves is the nucleation rate (see Table 2). The x axis shows the time between the start and finish of nucleation.

Table 2. Nucleation rates for different cooling rates in the quench process calculated from the fitted lines in Fig. 12.

| Cooling rate ($10^{11}$ Ks$^{-1}$) | 58.30 | 5.83 | 0.58 |
|---|---|---|---|
| Nucleation rate ($I$) ($10^{35}$ m$^{-3}$s$^{-1}$) | 1.12 | 0.41 | 0.08 |

The isothermal simulations in the previous section showed that the nucleation rate is temperature dependent. In quenching crystallization begins by formation of small clusters of atoms at high temperatures. In section 3.5 (Fig. 10) the nucleation regimes for quenching show that the crystallization generally occurs between 586 K and 725 K. In a slower cooling rate, the crystallization occurs at a higher temperature.

In Fig. 13 we show that the nucleation rates in the quenching process and the isothermal cases with high annealing temperatures are almost similar. At the highest cooling rate of 5.83x $10^{12}$ Ks$^{-1}$ the nucleation rate is 1.12x$10^{35}$ m$^{-3}$s$^{-1}$ (Table 2) lies between the nucleation rates isothermal cases at annealing temperatures of 650 K (1.12x$10^{35}$ m$^{-3}$s$^{-1}$) and 700 K (0.07x$10^{35}$ m$^{-3}$s$^{-1}$). The rate of nucleation is calculated using the same procedure used for isothermal process. Nucleation rate at cooling rate of 5.83x $10^{10}$ Ks$^{-1}$ is very close to the nucleation of 725 K.

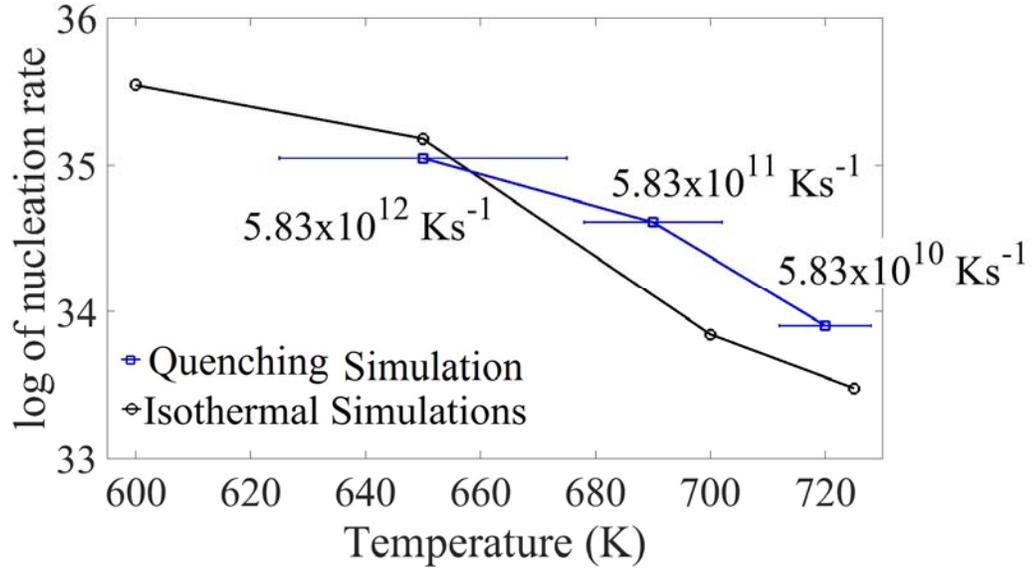

Fig. 13. The log of nucleation rates for isothermal (600-725 K) and quenching cases are plotted. The corresponding nucleation temperature range is shown for the quenching simulations by horizontal bars.

3.8 Comparison with classical nucleation theory

CNT can provide some insights on the homogeneous nucleation process. CNT suggests that there is a free (activation) energy barrier, $W^*$, for formation of a solid nucleus with a critical size of $r^*$. The nucleation typically happens when the probability of energy fluctuation is sufficient to overcome the activation barrier. The probability of energy fluctuation is given by the Arrhenius type equation and the rate of homogeneous nucleation is [53-56],

$$I = I_0 \exp\left(-\frac{W^*}{k_B T}\right), \tag{1}$$

where $T$ is the temperature, $k_B$ is the Boltzmann constant, and $I_0$ is a coefficient that depends on temperature and the interface free energy, $\sigma_{SL}$ [56]. $W^*$ is defined by [57],

$$W^* = \left( \frac{16\pi}{3} \frac{\sigma_{SL}^3}{(\Delta G_V)^2} \right), \qquad (2)$$

$\Delta G_V$ is the difference between the free energies of liquid and solid crystal per unit volume. If the change in molar heat capacities is constant, $\Delta G_V$ according to Hoffman is equal to $\Delta H_m \left( T \Delta T / T_m^2 \right)$ [58, 59], where $\Delta T$ is the undercooling ($\Delta T = T_m - T$), and $\Delta H_m$ is enthalpy of melting. By combining Eq. (1) and Eq. (2), the homogeneous nucleation rate becomes:

$$I = I_0 \exp\left[ \left( -\frac{16\pi \sigma_{SL}^3 T_m^4}{3 k_B (\Delta H_m)^2} \right) \frac{1}{T^3 (\Delta T)^2} \right] = I_0 \exp\left( -\frac{A}{T^3 (\Delta T)^2} \right), \qquad (3)$$

where, $A$ is a constant that depends on the solid-liquid interface energy and enthalpy. Eq. 3 also suggests that homogeneous nucleation rate strongly depends on the undercooling or the annealing temperature. The nucleation rate is maximum at the critical temperature. The critical temperature can be derived from Eq. 3 by setting its first derivative to zero. This suggests that the critical temperature is $T_{cr} = \frac{3 T_m}{5}$ (~550 K). As it was mentioned before, the calculated critical temperature from MD is $\sim \frac{T_m}{2}$ (475 K), which is a reasonable estimation from MD simulations and close to the CNT and experimental values of critical temperature of nucleation, which lies between 0.5-0.6 times of the melting temperature [60, 61].

We can also find the critical radius from CNT, which is suggested to be:

$$r^* = 2 \frac{\sigma_{SL}}{\Delta G_V}, \qquad (4)$$

We previously calculated $\sigma_{SL}$, the specific free energy of the critical nucleus formation is estimated to be the interface the solid-liquid interface free energy of 172.6 mJ-m$^{-2}$ and $\Delta H_m$ to be 11.50 kJmol$^{-1}$ for Al [23]. The atomic volume in solidification is available from isothermal simulation. By utilizing Eq. 2 and considering the normalized temperature for annealing, $T_{normalized} = T / T_m$, $\Delta G_V$ is calculated for different annealing temperatures. So according to CNT the calculated critical radius (size/diameter) lies between 1.25 (2.5) nm and 2.0 (4.0) nm for different annealing temperatures.

The prediction of critical size from CNT is dependent on the annealing temperature (Fig. 14). In section 3.2, we showed at that the critical size calculated by MD simulations is between ~0.82 nm and 4 nm in the isothermal cases, and it is between ~1.8 to ~4.5 nm for quenching cases. CNT predicts almost similar critical sizes to MD simulations from 650 K. But CNT estimates the critical size to be higher than the values obtained from both isothermal and quenching simulations for lower annealing temperatures. In MD simulations of quenching, the size of the critical nucleus increases as the quench rate decreases. As shown before in Section 3.3 with a lower quench rate the nucleation starts at higher temperature, and this results in formation of a larger size critical nucleus.

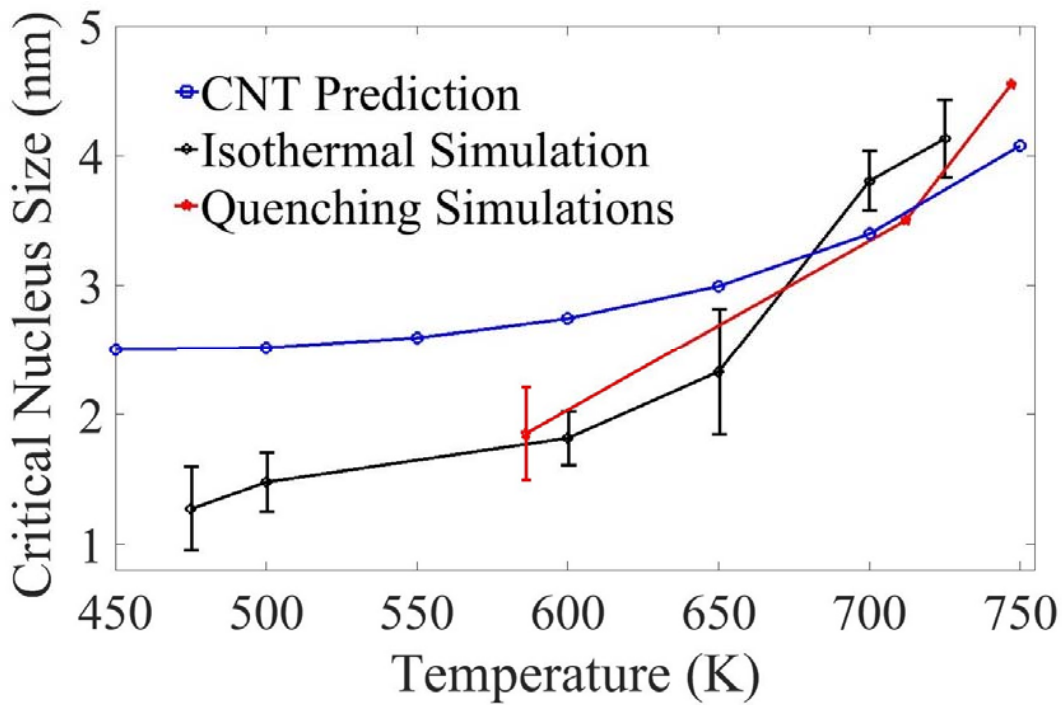

Fig. 14. The critical nucleus size calculated by CNT at different temperatures is compared with the results of the isothermal and quenching simulations.

CNT overestimate the critical nucleation size (Fig. 14) at low annealing temperatures (below 650 K). To calculate the critical nucleation size at different annealing (undercooling) temperatures by Eq. (4), $\sigma_{SL}$ at the melting point is used in our work as well as most of the other works in the literature [62-64]. This means the numerator of Eq. (4) is kept constant for calculating the critical size nucleus at different temperatures. However as it was demonstrated previously, $\sigma_{SL}$ decreases with lower annealing temperature (or increasing undercooling) [65, 66]. Therefore the numerator of Eq. (4) should also decrease with lower annealing temperature, making the critical size from CNT closer to MD simulation results.

3.9    Determination of induction time

In sections 3.6 and 3.7, nucleation rates are calculated for both isothermal and quenching cases which show how frequently nucleation events occur in the superheated melt of Al. For higher nucleation rates, a system can escape the metastable superheated liquid state and form the crystalline phase. The ability of a system to sustain small thermal fluctuations while in a metastable equilibrium state is characterized by the induction time, which is defined as the time elapsed between the establishment of supercooling and the appearance of persistent, stable nuclei [17]. The theory of homogenous nucleation suggests that the induction time is closely related to the nucleation rate, and the relationship depends on whether the system escapes the metastable state [17, 67, 68]. Nucleation can be divided into mono or polynuclear mechanisms [17]. When the system undergoes a phase transformation under conditions allowing the formation of many statistically independent nuclei it is called polynuclear mechanism, and for single nucleus it is called mononuclear mechanism. The formulations for the induction time for mononuclear, polynuclear, and combination of both mechanisms are given by Kashchiev et al. [68]. When the system volume is small, similar to our cases, polynuclear formulation reduces to that of the mononuclear case. The induction time for the mononuclear mechanism is given by,

$$\tau^* = \frac{1}{IV}, \tag{5}$$

where $V$ is the volume of the system and $I$ is the nucleation rate. Through this relationship, the induction time $\tau^*$ can be calculated from the previously obtained nucleation rate. It is worth mentioning that the role of $I$ is weaker in the polynuclear case than in the mononuclear case [17].

As $\tau^*$ refers to the time required for the system to escape from the metastable to a stable crystalline state, we can also assume that it is the minimum time required for the first crystalline nucleus to form.

Mullin [67] alternatively defined the induction time as $\tau^* = t_r + t_n + t_g$ ; the induction time is divided into three periods. $t_r$ is the relaxation time required for the system to achieve a quasi-steady-state distribution of molecules in the system; $t_n$ is the time required for the formation of the first stable nucleus (critical sized); and $t_g$ is the time between formation of the first stable nucleus (or nuclei) and the second stable nucleus (or nuclei) inside the melt, and after this time the cluster of crystalline atoms do not dissolve back into the liquid phase.

The definition of induction time is valid for the quenching cases. But for isothermal processes superheated melt is kept at an annealing temperature directly and the nucleation occurs immediately. The time difference between first and second critical nucleus is very small until 600 K. Only at higher annealing temperatures such as 700 K or 725 K, there is a detectable time between formation and growth of the first critical nucleus and the formation of a secondary nucleus. This is evident by comparing the snapshots of nuclei formation and growth during solidification for isothermal and quenching processes in Fig. 6(a) and (b). But as it was discussed before, in an isothermal process the whole process of crystallization happens without any change in temperature. It is not possible to generalize the induction time for isothermal processes, as we cannot get all the quantities for the Mullin's formulation for all the annealing temperatures. The isothermal process is equivalent to CNT which also assumes constant temperature for nucleation. Overall it is more meaningful to calculate the induction time for the quenching process.

During quenching solid atoms start gathering and attempt to form an initial nucleus before it reaches the critical size. The number of atoms and the size of the initial nucleus fluctuate for a few picoseconds before reaching the critical size. We refer to the time between the initial attempt to form a nucleus (20-25 clustered solid atoms) at a site and the formation of a critical size nucleus (1000-1500 clustered solid atoms, shown before in Fig. 3(b)) as the nucleus origin time ($t_o$). In Fig. 15, $t_o$ and $t_g$ are shown for the quench process at the cooling rate of 5.83x10$^{11}$ Ks$^{-1}$.

We first determined $t_o$, $t_n$ and $t_g$ for different cooling rates by utilizing snapshots of MD simulations. The induction times calculated by using Eq. (5) and the Mullin's definition [67] are presented in Table 3. The initial relaxation time for the melt at 1,325 K (150 ps) is not included in the reported induction times. The problem with calculating induction time from Mullin's original formula is related to $t_n$. $t_n$ is dependent on the superheat temperature and the nucleation rate. As it was shown previously in section 3.3, the first nucleus (nuclei) occurs between 586 K and 725 K for Al for different cooling rates, but $t_n$ will be significantly different for different cooling rates. In this work, the induction time is assumed to be the combination of $t_o$ and $t_g$. These two quantities must be minimum for the nucleation rate to be maximum and vice versa. The results show a pattern of gradually increasing induction time with slower cooling rate.

$t_o$ can be compared with the results of theory for $\tau^*$ in Eq. (5). However the theoretical values are much lower, because in theory, the induction time is based on the fact that nucleation is stationary [68]. Stationary means the temperature is constant throughout the solidification and a supersaturation is imposed on the system [68]. Stationary nucleation is the simplest case of nucleation. In a realistic nucleation system such as in quenching of a superheated melt, none of these conditions hold true.

Table 3. Induction time (ps) at different cooling rates.

| Cooling Rate ($10^{11}$ Ks$^{-1}$) | 58.3 | 5.83 | 0.58 |
|---|---|---|---|
| $t_n$ (MD) | 273.0 | 1,053.0 | 10,471.0 |
| $t_o$ (MD) | 5.0 | 12.0 | 15 |
| $t_g$ (MD) | 13.0 | 20.5 | 27.5 |
| $\tau^*(t_o+t_g)$ (Our Definition) | 18.0 | 32.5 | 42.5 |
| $\tau^*$ (Eq. (5)) | 0.57 | 1.46 | 7.52 |
| $\tau^*(t_n+t_g)$ (Mullin's Definition) | 286.0 | 1,073.5 | 10,498.5 |

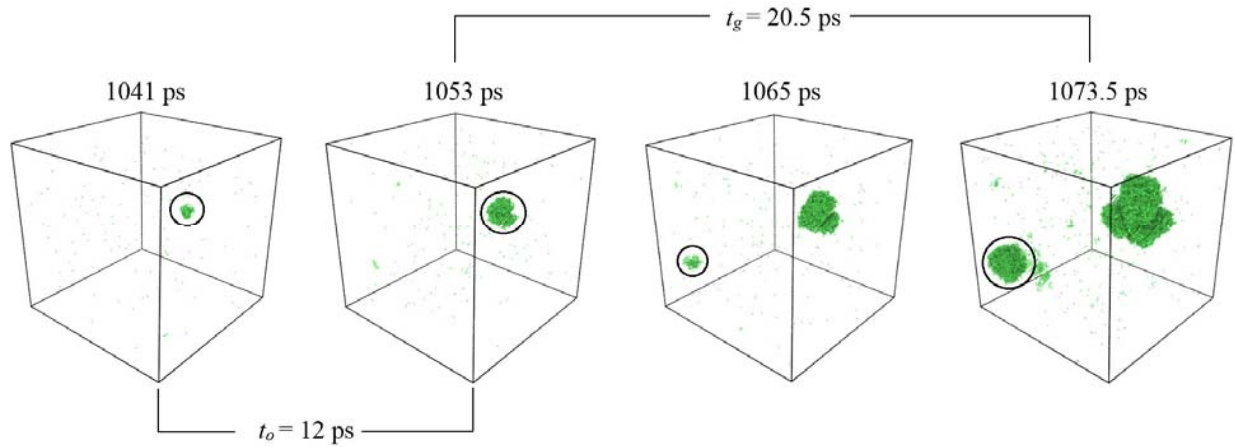

Fig. 15. Formation of first and second critical sized nuclei is shown for quenching at the cooling rate of $5.83 \times 10^{11}$ Ks$^{-1}$. The black circles show the process of first and second nuclei formation. Few solid atoms made the first attempt to form a solid cluster at 1,041 ps. At 1,053 ps, the first critical sized nucleus of 3.9 nm diameter is observed. Stable first nucleus and unstable second nucleus is shown at 1,065 ps. Second nucleus reaches the critical size at 1,073.5 ps.

3.10  Grain growth and microstructural evolution

Grain growth is usually defined as an increase in the mean grain size in polycrystals with an increase in annealing time. As discussed in Section 3.5, solid-state grain growth occurs as soon as the simulation box is completely solid (the third regime), and this phenomenon is interesting both from the experimental and theoretical points of view, as it affects the mechanical properties of materials.

To study solid-state grain growth, the simulation box is quenched from 1,325 K to 450 K, and then the resulting nanostructure is annealed at temperatures between 300 K and 725 K for 3,000 ps. The average grain size before starting the annealing process was ~5 nm.

Insignificant grain growth is observed for annealing temperatures lower than 450 K, (such as at 400 K and lower in Fig. 16(a)). At higher annealing temperatures, the grain boundary motion results in formation of larger grains (Fig. 16(b) and Fig. 16(c)). The effect of temperature on grain growth is related to the mobility of atoms. This is also very relevant to experimental observations where more grain growth is generally detected at higher annealing temperatures [69-72].

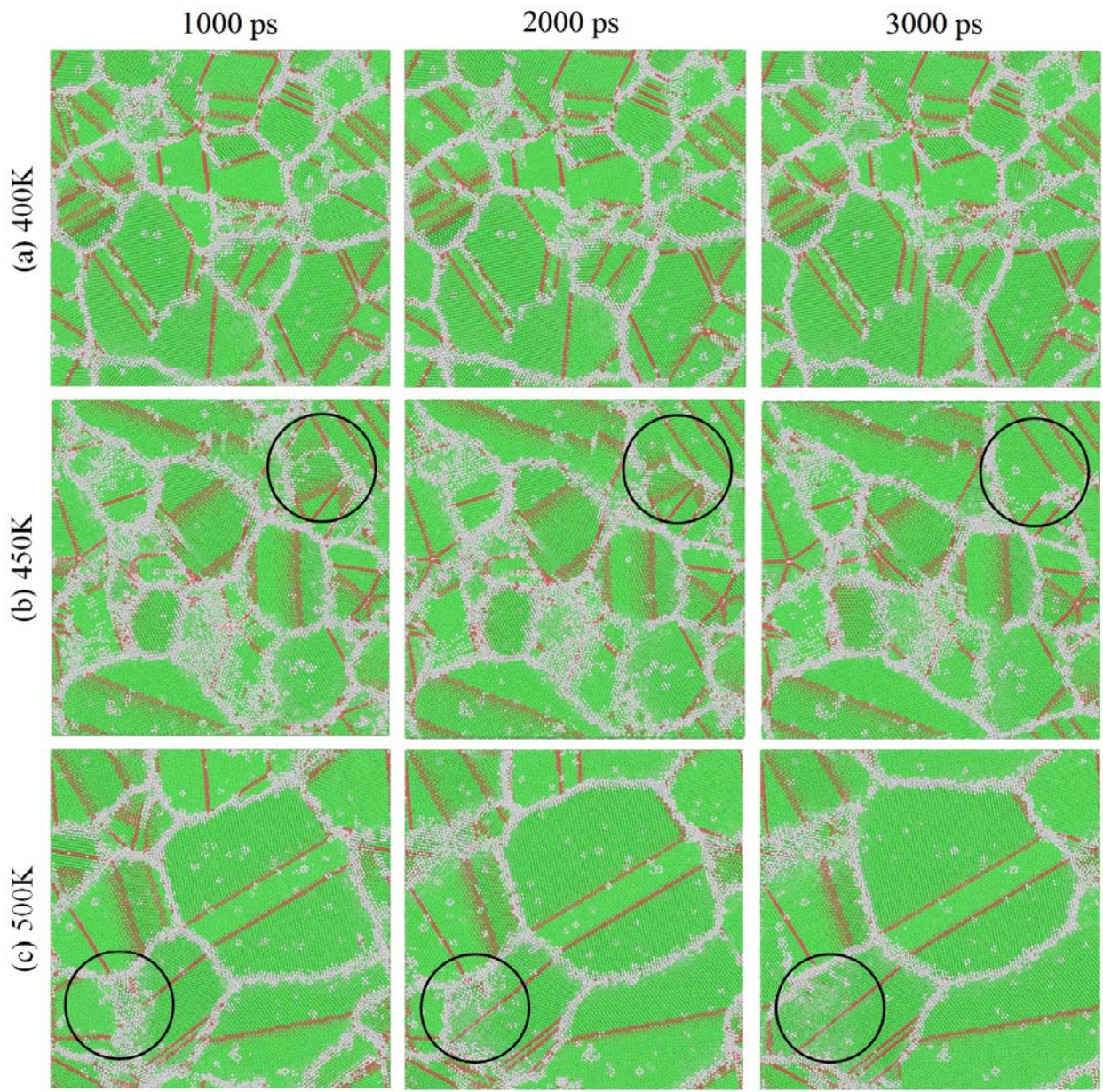

Fig. 16. Snapshots of a 20 nm by 20 nm cross section from the simulation box. The simulation box was quenched from 1,325 K to 450 K, and then the resulting nanostructure was annealed at (a) 400 K, (b) 450 K and (c) 500 K. The black circles in (b) and (c) show the area where the grain growth happens and fcc atoms replaced amorphous solid atoms.

The average grain size versus simulation time is shown for different temperatures in Fig. 17. The grain growth starts immediately for annealing temperatures higher than 600 K. For annealing temperatures below 450 K, the grain size remains below 10 nm at the end of 3,000 ps of annealing, whereas at 600 K the grains become as large as 15-20 nm. At 600 K and higher annealing temperatures no separate grains remains at the end of 3,000 ps of annealing, and the simulation box turned into a large single crystal, with a few stacking faults.

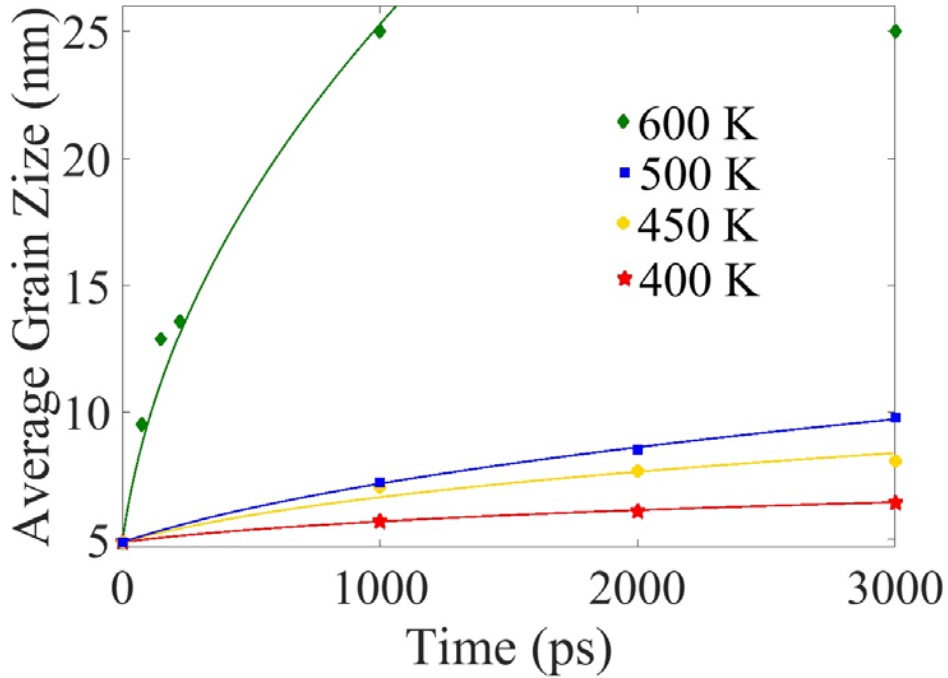

Fig. 17. Average grain size versus simulation time at different annealing temperatures. Each data point is the average of five different simulations at the specific annealing temperature. The dashed lines show the result of the fit using Eq. (6) with parameters given in Table 4.

The temperature dependent grain growth can also be explained using a grain growth exponent ($n$). Grain growth can be described by a power law [73-75],

$$(D/D_0)^{1/n} - 1 = Kt,  \qquad (6)$$

where $D_0$ is the initial average grain size before annealing (at $t=0$), $D$ is the average grain size after a period of annealing, $t$ is the time, and $K$ is the overall rate constant. $n$ is the grain growth exponent which depends on various factors such as grain boundary area, surface area, grain volume, and number of grains. The parameters are determined by fitting Eq. 6 to the simulation data (see Fig. 17) and the results for $n$ and $K$ are presented in Table 4.

From Table 4, the grain growth exponent remains less than the ideal value (0.5 for parabolic growth). At 600 K the growth is almost parabolic until ~1,000 ps. After 1,000 ps the simulation box becomes a single crystal and the model does not apply. For lower annealing temperatures the smaller values of $n$ signify slower grain growth.

Table 4. The grain growth parameters $n$ and $K$ (ps$^{-1}$) in Eq. (6) for Al at various annealing temperatures.

|  | Temperature (K) | | | |
| --- | --- | --- | --- | --- |
| Exponent | 400 K | 450 K | 500 K | 600 K |
| $n$ | 0.15 | 0.27 | 0.36 | 0.47 |
| $K$ | 1.77x10$^{-3}$ | 2.11 x10$^{-3}$ | 3.1 x10$^{-3}$ | 3.18 x10$^{-2}$ |

## 4. Conclusions

Homogenous nucleation from Al melt was investigated by million-atom MEAM-MD simulations. The main challenge of experimental studies of homogenous nucleation from pure Al is to observe the formation and growth of nuclei inside the melt during the solidification period, and the current work has enabled overcoming this challenge. We used both visual analysis such as direct observation of nuclei, and quantitative analysis of the data such as nucleation rate, induction time, fcc/hcp volume fraction, etc., to study the homogeneous nucleation process. Our MD simulations of homogenous nucleation utilizing a 3D simulation box with maximum of 5 million time steps allowed investigating the isothermal solidification process for 0.5 nanosecond and the quenching solidification process up to 15 nanoseconds.

Inspections by CNA showed that each nucleus had mainly fcc atoms with some hcp atoms. As the solidification process progressed, the hcp crystalline atoms aligned themselves to form stacking faults.

The average size of critical nuclei was determined to be between ~0.82 nm and ~4 nm in the isothermal processes, and between 1.8 nm and 4.5 nm in the quench processes. The size of critical nuclei follows the predictions of CNT. In the isothermal processes with annealing temperatures between 300 K to 475 K the critical nucleus size doesn't change significantly. But after that till 725 K the critical size increases with increasing annealing temperature. A relatively large number of nuclei formed (>50 nuclei in 25 nm$^3$). Below 350 K, the nucleation phenomenon was suppressed by fast solidification due to a very high driving force of solidification, and in cooperation with a low mobility of atoms resulted in formation of more amorphous solid atoms and lowering the number of crystalline nuclei (<40 nuclei in 25 nm$^3$). Above 700 K, the number of critical nuclei was reduced (<10 nuclei in 25 nm$^3$); at these high temperatures since there is not nucleation and growth of considerable number of crystalline nuclei or amorphous solid atoms, the few crystalline nuclei can growth to a much larger size before the simulation box is completely solid.

Utilizing the potential energy and percent crystalline atoms versus temperature data for quenching simulations (Fig. 10(c) and Fig. 10(d)), the solidification process can be divided into three temperature based thermodynamics regimes, where the specific temperatures ($T_{sc}$ and $T_{gg}$) depend upon the quench rate:

    i.    Sub-critical unstable nucleation regime above $T_{sc}$,

    ii.    Super-critical Stable nucleation regime between $T_{sc}$ and $T_{gg}$, and

    iii.    Solid-state grain growth regime below $T_{gg}$.

These regions were not clearly seen for isothermal cases with low annealing temperatures. Only at high temperature annealing of 650 K, 700 K and 725 K, could these three distinct regions be observed. The change in instantaneous temperature during nucleation (i.e. solidification) indicated that quenching is more realistic simulation procedure to study a nucleation process. As cooling rate decreases, the $T_{sc}$ moves towards the melting point.

We also determined the percentage of different type atoms for both isothermal and quenching cases. In the isothermal cases with higher annealing temperatures such as 700 K and 725 K, the percentage of fcc atoms (~60-65 %) was higher compared with that of the cases with lower annealing temperatures (~50-55 %). At very low annealing temperatures such as 300K and 350 K, the percentage of fcc atoms was very low (< 45%). In the quenching cases, by decreasing the cooling rate from 5.83x $10^{12}$ Ks$^{-1}$ to 5.83x $10^{10}$ Ks, the percentage of fcc atoms increased from ~20% to ~80%.

To determine the critical temperature for homogenous nucleation in the isothermal cases, the nucleation rate was calculated by plotting the number of nuclei versus time. The critical temperature of Al was determined to be ~475 K, with a maximum nucleation rate of $5.74 \times 10^{35}$ m$^{-3}$s$^{-1}$. The nucleation rate in quenching simulations was determined to be one to two orders of magnitude lower than that in isothermal cases with annealing temperatures lower than 600 K. This was attributed to the fact that in the quenching cases the nucleation occurred only between ~747 K to ~586 K, however in the isothermal cases with low annealing temperatures the nucleation and solidification occurred almost instantly. The nucleation rates for the isothermal cases with annealing temperatures of 700 K and 725 K are almost the same as those for quenching cases. Since nucleation during quenching occurs at much higher temperature than the critical temperature, it is not clear that the critical temperature and maximum nucleation rate has any significance for the actual nucleation process.

The critical nucleus size and the critical temperature for nucleation determined by MD simulations were compared to the CNT predictions. The critical temperature for nucleation obtained from CNT was close to the results obtained by MD simulations for the isothermal cases. The calculated critical size of nucleus using CNT increases with increasing annealing temperature, and is very close to the values obtained from MD simulations above 650 K. But, CNT estimates the critical size to be higher than MD simulations for lower annealing temperatures, and this is because we have assumed $\sigma_{SL}$ is independent of temperature. Since the solid-liquid interface energy is expected to decrease with decreasing temperature, using a temperature dependent $\sigma_{SL}$ will result in an additional decrease in the critical size of the nucleus at lower temperatures, confirming the MD simulation results.

The induction time, which is closely related to the nucleation rate, was also calculated by MD simulation results. In theory (Eq. (5)), the induction time is inversely related to the nucleation rate and gives the time for formation of the first critical nucleus; however since it assumes a constant temperature and superheated melt throughout the solidification, it does not reasonably mimic the realistic experimental conditions. We compared the theoretical value of induction time to our defined nucleus origin time ($t_o$), both showing it increased by decreasing the cooling rate. We defined the actual induction time to be the time from the initial stages (3-5 clustered crystalline atoms) of formation of the first critical nucleus (nuclei) until the formation of the second critical nucleus (nuclei) ($t_o + t_g$).

Significant grain growth occurred in a temperature region above 500 K and below 650 K. At lower annealing temperatures, low mobility of atoms results in a very low grain growth rate. Grain growth exponent (*n*) increased by increasing the annealing temperature, and it reached the ideal value of 0.5 at 600 K.

## Acknowledgments

Authors would like to acknowledge the funding support from the National Science Foundation under Grant No. NSF-CMMI 1537170. The authors are grateful for computer time allocation provided by the Extreme Science and Engineering Discovery Environment (XSEDE), award number TG-DMR140008.